\documentclass[
prd,twocolumn
,superscriptaddress,noshowpacs,nofootinbib
,tightenlines
]{revtex4-1}
\usepackage{mathrsfs}
\usepackage{latexsym,bm}
\usepackage{graphicx}
\usepackage{indentfirst}
\usepackage{slashed}
\usepackage{amssymb}
\usepackage{amsmath}
\usepackage{bbm}
\usepackage{color}
\usepackage[dvipsnames]{xcolor}
\usepackage{epsfig}
\usepackage[titletoc]{appendix}
\usepackage{multirow}%
\usepackage{rotating}
\usepackage{epstopdf}
\usepackage{extarrows}
\usepackage{tabularx}
\usepackage{soul} 
\usepackage{xcolor}
\usepackage{lipsum}
\usepackage[colorlinks=true,allcolors=BlueViolet,hyperfootnotes=false]{hyperref}

\usepackage[utf8]{inputenc}

\newcommand{\be}{\begin{equation}} \newcommand{\ee}{\end{equation}}
\newcommand{\ba}{\begin{array}{c}} \newcommand{\ea}{\end{array}}
\newcommand{\bea}{\begin{eqnarray}} \newcommand{\eea}{\end{eqnarray}}

\newcommand{\A}{{\cal A}}

\begin{document}
\title{\Large New parametrization of the form factors in $\bar{B}\to D\ell\bar{\nu}_\ell$ decays}

\author{De-Liang~Yao}
\affiliation{School of Physics and Electronics, Hunan University, Changsha 410082, China}
\affiliation{Instituto de F\'{\i}sica Corpuscular (centro mixto CSIC-UV), Institutos de Investigaci\'on de Paterna,
Apartado 22085, 46071, Valencia, Spain}

\author{Pedro~Fernandez-Soler}
\affiliation{Instituto de F\'{\i}sica Corpuscular (centro mixto CSIC-UV), Institutos de Investigaci\'on de Paterna,
Apartado 22085, 46071, Valencia, Spain}

\author{Feng-Kun Guo}
\affiliation{CAS Key Laboratory of Theoretical Physics, Institute of Theoretical Physics, Chinese Academy of Sciences, Beijing 100190, China}
\affiliation{School of Physical Sciences, University of Chinese Academy of Sciences, Beijing 100049, China}

\author{Juan~Nieves}
\affiliation{Instituto de F\'{\i}sica Corpuscular (centro mixto CSIC-UV), Institutos de Investigaci\'on de Paterna,
Apartado 22085, 46071, Valencia, Spain}

\date{\today}
\begin{abstract}
A new model-independent parametrization is proposed for the hadronic form factors in the semileptonic $\bar{B}\to D\ell\bar{\nu}_\ell$ decay. By a combined consideration of the recent experimental and lattice QCD data, we determine precisely the Cabibbo-Kobayashi-Maskawa matrix element $|V_{cb}|=41.01(75)\times 10^{-3}$  and the ratio $\mathcal{R}_D=\frac{\mathcal{BR}(\bar{B}\to D \tau \bar{\nu}_\tau)}{\mathcal{BR}(\bar{B}\to D \ell \bar{\nu}_\ell)}=0.301(5)$. The coefficients in this parametrization, related to phase shifts by sumrulelike dispersion relations and hence called phase moments, encode important scattering information of the $\bar{B}\bar{D}$ interactions which are poorly known so far. 
Thus, we give strong hints about the existence of at least one bound and one virtual $\bar B \bar D$ $S$-wave $0^+$ states, subject to uncertainties produced by potentially sizable inelastic effects. This formalism is also applicable for any other semileptonic processes induced by the weak $b\to c$ transition.

\end{abstract}
%

\maketitle

%
%
\section{Introduction}
%
%
One of the most primary goals in flavor physics currently is to
precisely determine the elements of the
Cabibbo--Kobayashi--Maskawa~(CKM) matrix,
since they afford a sharp probe of physics beyond the standard
model (SM) as inputs of the CKM unitarity triangle.
For that purpose, experimental and theoretical efforts are
extensively devoted to study both inclusive and exclusive
semileptonic decays of bottom hadrons. 
For the latter ones, different ways have been 
proposed to parametrize the hadronic form factors involved,
the most commonly used of which are the Boyd-Grinstein-Lebed
(BGL)~\cite{Boyd:1997kz} and Caprini-Lellouch-Neubert
(CLN)~\cite{Caprini:1997mu} parametrizations.
There is tension in the determination of some of the entries
like $V_{cb}$ from $B$ meson decays, for which the result
considering inclusive decays~\cite{Alberti:2014yda}  
is larger than the value obtained from exclusive ones---a discrepancy at
$2$-$3\sigma$ significance level has existed since $2015$;
see \textit{e.g.}~Refs.~\cite{Aoki:2016frl,Amhis:2016xyh,Koppenburg:2017mad,Ricciardi:2018bgs} for recent reviews.
The main source of exclusive $V_{cb}$ determinations is the 
$\bar B\to D^{(*)}\ell \nu$ semileptonic decay.

Since 2015, significant progress has been made.
The Belle Collaboration measured  the differential decay rates of
the exclusive 
$\bar{B}\to D\ell\bar{\nu}_\ell$~\cite{Glattauer:2015teq}
and
$\bar{B}\to D^*\ell\bar{\nu}_\ell$ reactions~\cite{Abdesselam:2018nnh}
using their full data set;
and there have been lattice QCD (LQCD) results on the form factors
at nonzero recoils for $\bar{B}\to D\ell\bar{\nu}_\ell$ 
obtained by the HPQCD~\cite{Na:2015kha} and Fermilab Lattice plus
MILC (FL-MILC)~\cite{Lattice:2015rga} Collaborations.
It turns out that the CLN and BGL parametrizations lead to different values of the extracted $|V_{cb}|$, 
see \textit{e.g.},
Refs~\cite{Glattauer:2015teq,Bigi:2016mdz,Jaiswal:2017rve,Abdesselam:2018nnh}.
For instance, the Belle determinations of this CKM matrix element from the $B\to \bar D \bar \ell \nu$ decay  are $(39.86\pm1.33)\times10^{-3}$ or $(40.83\pm1.13)\times10^{-3}$ using the CLN or BGL parametrizations, respectively ~\cite{Glattauer:2015teq}. For comparison, the updated HFLAV averages for the inclusive determination of  $|V_{cb}|_\text{in}$ are $(42.19\pm 0.78)\times 10^{-3}$ or $(41.98\pm 0.45)\times 10^{-3}$ depending on the used scheme~\cite{Amhis:2016xyh}.
It is pointed out in Refs.~\cite{Bigi:2016mdz,Bigi:2017njr,Grinstein:2017nlq} that the CLN parametrization, based upon heavy quark effective theory, though very useful in the past, may no longer be adequate to cope with the accuracy of the currently available data. The BGL parametrization is a model-independent expansion in powers of a small variable $z$. To describe data, the expansion needs to be truncated at least at the $z^2$ order, leading to three unknown coefficients for each form factor. The relation $f^+=f^0$ at $q^2=0$ imposes a constraint among these parameters, which on the other hand do not have an obvious physical interpretation, except  for those of the leading term that could be related to the form factor normalization.

In this article, we propose a new model-independent parametrization based on a dispersion relation, {which successfully describes  high-accuracy Belle and LQCD data and allows for an extraction of $|V_{cb}|$ with a small uncertainty comparable to that obtained from BGL fits. Furthermore, all the involved parameters are physically meaningful, encoding  scattering information on elastic and inelastic $\bar{B}\bar{D}$ interaction through dispersion relations to phase shifts. Having  validated the new parametrization, the crucial point resides in the physical meaning that can be ascribed to the fitted parameters, which is however, difficult to find for the BGL and CLN ones.   

The interaction between two heavy-light mesons close to threshold has attracted much attention in recent years triggered by the discoveries by BaBar, Belle, BESIII, LHCb and other experiments of many exotic hidden charm/bottom states, like the isoscalar $X(3872)$  or isovector $Z_c(3900)$, $Z_b(10610)$ or $Z_b(10650)$ resonances, that cannot be accommodated as  $\bar c c$ or $\bar b b$ (see e.g. general discussions in Refs.~\cite{Cerri:2018ypt} and \cite{Guo:2017jvc}). Actually, they are believed to be largely hadronic molecules, i.e.,  $H^{(*)} \bar H^{(*)}$ loosely bound or resonant states ($H=B$ or $D$). This possibility opens  new exciting scenarios  to learn about details of the nonperturbative QCD regime. The hidden charm and bottom spectra should have  a counterpart with $b\bar c$ heavy-flavor content, and the scheme established in this work offers an interesting opportunity to obtain some model-independent constraints from the existing accurate data on semileptonic decays.}

The $\bar{B}\bar{D}$ interaction, related to the $\bar{B}\to D$ transition amplitude by crossing, is poorly known so far; however, it is essential to explore the spectrum of  hadrons containing one bottom quark  and one charm antiquark, i.e., $B_c$ mesons; see Ref.~\cite{Sakai:2017avl} for example. Up to now, the discovery of the $B_c$ mesons is restricted to two states only~\cite{Tanabashi:2018oca}: $B_c(6275)$ and $B_c(2S)(6871)$, both with $J^P=0^-$ (although the vector $B_c^*(2S)$ was reported recently by both ATLAS~\cite{Sirunyan:2019osb} and LHCb~\cite{Aaij:2019ldo}, its mass has not been measured because of the unconstructed low-energy photon in both experiments). In view of the well established bottomonium or charmonium spectra, it is clear that many $B_c$ states are still missing. Hopefully,  states will be unraveled in the near future due to the advent of the LHCb, which is an efficient factory to produce $b\bar{c}$ or $bc$ states. Besides, prognosis of charmed-bottom hadrons from LQCD has been made very recently ~\cite{Mathur:2018epb}. Our new parametrization, bringing information from semileptonic decays to the scattering problem, will definitely shed light on those newly predicted/discovered states.

%
%
%

\section{New parametrization}

To proceed, let us first introduce  the semileptonic $
\bar{B}(p)\to D(p^\prime) \ell(q_1) \bar{\nu}_\ell(q_2)$ differential decay rate~\cite{Korner:1989qb}
\bea\label{eq.dgdq2}
\frac{{\rm d}\Gamma}{{\rm d} q^2}=\frac{8\mathcal{N}|\vec{p}^{\,\ast}|}{3}\bigg[\bigg(1+\frac{m_\ell^2}{2q^2}\bigg)|H_0|^2+\frac{3m_\ell^2}{2q^2}|H_t|^2\bigg]\ ,
\eea
with $q\equiv p-p^\prime=q_1+q_2$ and $|\vec{p}^{\,\ast}|$ being the
modulus of the three-momentum of the $D$ meson in the $\bar{B}$ rest frame.
The normalization factor is
\bea
\mathcal{N}=\frac{G_F^2}{256\pi^3}\eta_{\rm EW}^2|V_{cb}|^2\frac{q^2}{m_B^2}\bigg(1-\frac{m_\ell^2}{q^2}\bigg)^2\ ,
\eea
where $G_F=1.166\times 10^{-5}$~GeV$^{-2}$ is the Fermi coupling constant and the factor $\eta_{\rm EW}=1.0066$ accounts for the leading order electroweak corrections~\cite{Sirlin:1981ie}. Here $m_B$ ($m_D$) and $m_\ell$ denote the masses of the $B$ ($D$) meson and the lepton, respectively. We use the values $m_D=1867.22$~MeV, $m_B=5279.47$~MeV and $m_\tau=1776.91$~MeV.
Furthermore, the helicity amplitude $H_0$ amounts to the longitudinal part of the spin-1 hadronic contribution, while $H_t$ corresponds to the spin-0 hadronic contribution, owing its presence to the off-shellness of the weak current. They are related to the conventional hadronic vector ($J^P=1^-$) and scalar ($J^P=0^+$) form factors, i.e., $f_+(q^2)$ and $f_0(q^2)$, respectively, through
\bea
H_0=\frac{2m_B|\vec{p}^{\,\ast}|}{\sqrt{q^2}}f_+(q^2) \ , \quad H_t=\frac{m_B^2-m_D^2}{\sqrt{q^2}}f_0(q^2) \ .
 \eea
At $q^2=0$, the two form factors coincide: $f_+(0)=f_0(0)$.

According to Refs.~\cite{Lepage:1979zb,Lepage:1980fj,Brodsky:1981rp}, and using general arguments from QCD, one expects vector and scalar form factors to fall off as $1/s$ (up to logarithms) when $|s|\to \infty$. Thus, based on analyticity, unitarity and crossing symmetry,  once-subtracted dispersion relations for each form factor admit solutions of the Omn\`es form 
\bea\label{eq.fpf0para}
f_{i}(q^2) = f_{i}(s_0)\exp\bigg[\frac{q^2-s_0}{\pi}\int_{s_{\rm th}}^\infty\frac{{\rm d} s}{s-s_0}\frac{\alpha^i(s)}{s-q^2}\bigg],
\eea
for $q^2<s_{\rm th}$. 
In addition,  $i=+,0$, $s_{\rm th}=(m_B+m_D)^2$ is the $\bar B\bar D$  threshold, $s_0$ is the subtraction point and $\alpha^i(s)$ is the phase of the corresponding form factor. 
This solution can be easily obtained noticing {$f_i(s+i\epsilon) = |f_i(s)|e^{i\alpha_i(s)}$ and using the Schwarz reflection principle $f_i(s-i\epsilon) = f_i^*(s+i\epsilon)  = |f_i(s)|e^{-i\alpha_i(s)}$  so that $\ln f_i(s+i\epsilon) - \ln f_i(s-i\epsilon) = 2i\alpha^i(s)\, \theta(s-s_\text{th})$.} It is worthwhile to emphasize that Eq.~\eqref{eq.fpf0para} holds even in the inelastic regime, i.e., when channels with a higher threshold such as $\bar B^* \bar D^*$ are open. In the elastic region ($\sqrt{s}<m_{B^*}+m_{D}$ for $i=+$ and $\sqrt{s}<m_{B^*}+m_{D^*}$ for $i=0$), the phase $\alpha^i(s)$ coincides with the $P$- and $S$-wave $\bar B\bar D$ scattering phase shift for $f^+$and $f^0$, respectively, according to the Watson's theorem~\cite{Watson:1952ji}.

In the physical $\bar B\to D\ell\bar \nu_\ell$ decay, the maximum value of $q^2$ is $q_\text{max}^2=(m_B-m_D)^2$.
Given that $s \ge s_{\rm th} \gg q_{\rm max}^2\ge q^2$,
Eq.~\eqref{eq.fpf0para} can be recast into a new form,
\bea
f_{i}(q^2) = f_{i}(s_0)\prod_{n=0}^{\infty}\exp\bigg[\frac{q^2-s_0}{s_{\rm th}}\A_n^i\frac{q^{2n}}{s_{\rm th}^n}\bigg]\ ,
\label{eq.new}
\eea
with the dimensionless coefficients (phase moments) defined as
\bea\label{eq.moments}
\A_n^i\equiv\frac{1}{\pi}\int_{s_{\rm th}}^{+\infty}\frac{{\rm d}s}{s-s_0}\frac{\alpha^i(s)}{(s/s_{\rm th})^{n+1}}\ .
\eea
Since the power of $s$ in the denominator of the integrand above grows as $n+1$, higher moments become sensitive only to the details of the form-factor phases $\alpha^i(s)$ in the vicinity of threshold. Equation~\eqref{eq.new} provides a new parametrization of the form factors in $\bar{B}\to D$ semileptonic decays. The coefficients $\A_n^i$ are called {\it phase moments} hereafter, due to the fact that they are related to the phases of the form factors in the physical  $\bar{B}\bar{D}$ scattering region.

{Note that the exponential form for form factors was also obtained within the Isgur-Scora-Grinstein-Wise constituent quark potential model through completely different reasoning~\cite{Isgur:1988gb, Grinstein:1986ad}.}

%
%

\section{Fit to Belle and LQCD data}

Let us first define the recoil variable $\omega = (m_B^2+m_D^2-q^2)/(2m_Bm_D)$. It ranges from 1 at zero recoil, $q^2 = (m_B-m_D)^2$, to about $(m_B^2+m_D^2)/(2m_Bm_D)\approx 1.59$ at $q^2=0$, for the decays into electron or muon leptons.  
To determine the phase moments $\A_n^{i}$ introduced in Eq.~\eqref{eq.new}, we perform a combined fit to the recent experimental data measured by  Belle~\cite{Glattauer:2015teq} together with the LQCD results of the vector and scalar form factors at nonzero recoil obtained by the HPQCD~\cite{Na:2015kha} and FL-MILC~\cite{Lattice:2015rga} collaborations. 

The Belle data consist of the weighted averaged differential decay rates for 10 $\omega$-bins (see Table~II of Ref.~\cite{Glattauer:2015teq}), and should be confronted with 
\bea
\frac{\Delta \Gamma_k}{\Delta\omega}=\frac{1}{\Delta\omega}\int_{\omega_{k,{\rm min}}}^{\omega_{k,{\rm max}}}\frac{{\rm d}\Gamma}{{\rm d} \omega}{\rm d} \omega\ ,\quad k=0,\cdots,9\ ,
\eea
where the $\Delta \omega$ is the width of each bin, $\omega_{k,{\rm min\, (max)}}$ is the minimal (maximal) value of $\omega$ in the $k$th bin. The lepton masses, except for the tau case
to be discussed later, are neglected.  

\begin{figure}[t]
\begin{center}
\epsfig{file=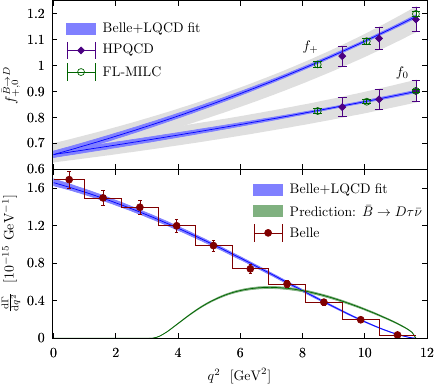,width=0.49\textwidth}
\caption{\label{fig.fit} Upper panel: vector and scalar form factors. Lower panel: differential decay rates. The gray bands stand for the HPQCD results from the BCL continuous parametrization provided in \cite{Na:2015kha}.}
\end{center}
\end{figure}

The FL-MILC Collaborations~\cite{Lattice:2015rga} provide results both for both $f_+$ and $f_0$ at three different $\omega\in\{1.00,1.08,1.16\}$. The HPQCD Collaboration~\cite{Na:2015kha} presents their results in terms of the Bourrelly-Caprini-Lellouch (BCL, a simple alternative to BGL, see Ref.~\cite{Bourrely:2008za}) parametrization for the entire kinematic decay region (see the gray bands in the upper panel of Fig.~\ref{fig.fit}). However, they only performed numerical lattice simulations for three different $q^2$ configurations, which lead to $\omega$ values  in the range of $[1,\sim 1.11]$. Therefore, as done in Refs.~\cite{Bigi:2016mdz,Jaiswal:2017rve}, we prefer to extract, from the BCL parametrization obtained in Ref.~\cite{Na:2015kha}, three values for each of the form factors, $f_+$ and $f_0$,  at $\omega\in\{1.00,1.06,1.12\}$. The 12 lattice data points with error bars are shown in the upper panel of Fig.~\ref{fig.fit}. We note that the HPQCD errors are significantly larger than the FL-MILC ones.

In our fit, in addition to the phase moments $\A_n^{i}$, the subtraction $f_0(s_0)$ and the CKM matrix element $|V_{cb}|$ are treated as free parameters as well. The kinematic constraint $f_+(0)=f_0(0)$ imposes a relation for the subtractions of both form factors
\bea
f_+(s_0) =f_0(s_0)\exp\bigg[\frac{s_0}{s_{\rm th}}(\A_0^+-\A_0^0)\bigg]\ .
\eea
We choose $s_0=0$ as the subtraction point, and  find that a truncation of the the expansion in Eq.~\eqref{eq.new} to the first order, i.e., $n=0$, is sufficient to accurately describe the data as seen in Fig.~\ref{fig.fit}. Consequently, we have a total of  four free parameters: $f_0(0)$, $\A_0^0$, $\A_0^+$ and $|V_{cb}|$. Fit results are collected in Table~\ref{tab:fit.combined}, where the errors in brackets are obtained from the minimization procedure. Moreover, it is found that the precision of the data set at hand is not sufficient to reliably pin down the phase moments $\A_n^i$ with $n\ge 1$. We already observe large correlation in Table~\ref{tab:fit.combined}.
\begin{table}[t]
\caption{Results from the combined fit to Belle~\cite{Glattauer:2015teq} and LQCD~\cite{Lattice:2015rga,Na:2015kha} data.}\label{tab:fit.combined}
\begin{tabular}{cl|rrrr}
\hline\hline
&  & \multicolumn{4}{c}{Correlation matrix}\\
\multicolumn{2}{c|}{$\frac{\chi^2}{dof} =\frac{6.47}{22-4}\simeq 0.36$} & $f_0(0)$ & $\A_0^0$   & $\A_0^+$  &$|V_{cb}|\times 10^3$ \\
\hline
$f_0(0)$           & $  0.658(17)$                 & $ 1.000$ & $ -0.979$ & $ -0.978$ & $-0.818$ \\
$\A_0^0$           & $  1.38(12)$                  & $ -0.979$ & $ 1.000$ & $ 0.957$ & $0.801$ \\
$\A_0^+$           & $  2.60(12)$              & $ -0.978$ & $ 0.957$ & $ 1.000$ & $0.774$ \\
$|V_{cb}|\times 10^3$ & $  41.01(75)$                 & $-0.818$ & $0.801$ & $0.774$ & $ 1.000$ \\
\hline\hline
\end{tabular}
\end{table}
In Fig.~\ref{fig.fit}, the form factors and the differential decay rates from the combined fit are plotted as a function of $q^2$ in the whole kinematic region. We also show the prediction of the differential decay rate for the $\bar{B}\to D\tau\bar{\nu}_\tau$ decay. For comparison, the Belle and LQCD (HPQCD and FL-MILC) data are displayed as well.

From the best fit, we get 
\bea
|V_{cb}|=(41.01\pm0.75)\times 10^{-3}\ ,
\eea
which is in agreement with the determination reported in Ref.~\cite{Glattauer:2015teq} using the BGL parametrization, but higher than the values obtained using the CLN one~\cite{Glattauer:2015teq,Bigi:2016mdz,Bigi:2017njr,Grinstein:2017nlq,Abdesselam:2018nnh}. It also agrees with the world average of the inclusive determinations~\cite{Amhis:2016xyh}. Our result confirms the conclusion that the previous tension between the exclusive and inclusive determinations was mostly due to the use of the CLN parametrization. The error in our determination is { 2\%, which is comparable to that obtained from} the combined fit in Ref.~\cite{Bigi:2016mdz} to the experimental data (BaBar~\cite{Aubert:2009ac} and Belle~\cite{Glattauer:2015teq}) and LQCD results (HPQCD~\cite{Na:2015kha} and FL-MILC~\cite{Lattice:2015rga}) using the BGL parametrization. Furthermore, as already commented, the fitted phase moments $\A_0^{0,+}$ provide valuable information to constrain  the $\bar B D$ interaction.  

Note that we  expect a systematic error  of the order of $q^2/s_{\rm th}$, which is negligible in the low $q^2$ region where there are no lattice results, and at most $q_{\rm max}^2/s_{\rm th}$ in the high $q^2$ region. This source of systematical uncertainties can be arbitrarily reduced by increasing the truncation order.

With the parameters in Table~\ref{tab:fit.combined}, we predict the ratio 
\bea\label{eq.RD}
\mathcal{R}_D=\frac{\mathcal{BR}(\bar{B}\to D \tau \bar{\nu}_\tau)}{\mathcal{BR}(\bar{B}\to D \ell \bar{\nu}_\ell)}=0.301(5)\, , 
\eea
with $\ell=e$ or $\mu$. It is well consistent with the predictions using the LQCD form factors: $\mathcal{R}_D=0.299(11)$ by FL-MILC~\cite{Lattice:2015rga} and $\mathcal{R}_D=0.300(8)$ by HPQCD~\cite{Na:2015kha}. However, the central value is significantly smaller than the values measured  by BABAR, $\mathcal{R}_D=0.440(58)(42)$~\cite{Lees:2012xj}, and by Belle, $\mathcal{R}_D=0.375(64)(26)$~\cite{Huschle:2015rga}. Yet, the deviation is $1.8~\sigma$ from the former and only $1.0~\sigma$  from the latter, given the large uncertainties in the experimental measurements. It is intriguing to see whether the deviation persists under more precise measurements. {Actually very recently, the Belle collaboration
announced a new preliminary measurement of $\mathcal{R}_D=0.307(37)(16)$~\cite{Abdesselam:2019dgh}  compatible with the SM at the $1.2\, \sigma$ level.}

We checked the dependence of the above results on the subtraction point $s_0$ by redoing fits with $s_0$ varied in the range $[0,q^2_{\rm max}]$. We find that the fit quality keeps exactly the same as for $s_0=0$, and the values of $|V_{cb}|$ and $\mathcal{R}_D$ are independent of the choice of $s_0$ as well. This is because in the Omn\`es representation, one is free to choose any $s_0$. The dependence of $s_0$ in the exponential in Eq.~\eqref{eq.fpf0para} or Eq.~\eqref{eq.new} is compensated by the parameter $f_0(s_0)$ that behaves as a normalization factor. 

{
The high  correlations between the fitted parameters could  weaken the statistical power of the fit in Table~\ref{tab:fit.combined}, leading to somehow unreliable and unstable estimates of the resulting observables. This is extremely important given the physical relevance that we ascribe here to  the phase moments. Thus, we have carried out a normality test by looking at the residual distribution central moments of order $r$, defined as:
\begin{equation}
    \mu_r= \frac{1}{n_D}\sum_{i=1}^{n_D} \left( \frac{R_i-\overline{R}}{\Delta R}\right)^r, \quad R_i= \frac{y_i^{\rm fit}-y_i^{\rm data}}{\Delta y_i^{\rm data}}
\end{equation}
with $R_i$, the resulting residuals of the fit. They should follow a normal distribution within a given confidence level (CL) \cite{Evans:2004}. We have $n_D=22$, the number of fitted inputs, $\overline R = \sum_{i=1}^{n_D} R_i/n_D$  and $(\Delta R)^2= \overline{R^2}-(\overline R)^2$, with $\overline{R^2}$ being the average of the $n_D$ values for $R_i^2$. Making use of the central limit theorem~\cite{Evans:2004}, implicitly accepting that $n_D$ is sufficiently large, it can be also shown that the variables $\mu_r$ are Gaussian distributed,  with means and standard deviations $\left(\langle\mu_r\rangle, \Delta\mu_r\right)$ collected in Table I of Ref.~\cite{Perez:2015pea} up to $r=8$\footnote{{We have also proceeded by Monte Carlo sampling $n_D$ Gaussian numbers and computing
the distribution of moments, as mentioned in the latter reference, to estimate deviations from the large $n_D$ limit predictions given in \cite{Perez:2015pea}, and found  very small corrections already for the $n_D=22$ case considered here.}}.  We have computed the estimators $|\mu_r^{\rm fit}-\langle\mu_r\rangle|/\Delta\mu_r$, and obtained $(0.12, 0.69, 0.02, 0.58, 0.00, 0.33)$ for $r=3, \cdots 8$. This shows that the fit of Table~\ref{tab:fit.combined} passes the normality test widely. Thus, the high correlations between the parameters do not lead to statistically unreliable or unstable estimates of the first-order phase moments  or $|V_{cb}|$, beyond the CL accounted for by the error bars quoted in Table~\ref{tab:fit.combined}.

In addition, we have also contemplated the possibility of performing a six-parameter fit, including the two higher moments ${\cal A}_1^{0,+}$ in the expansions of Eq.~\eqref{eq.new} for $f_+$ and $f_0$.  We find an improved description of the inputs, particularly of the FL-MILC LQCD results, with $\chi^2/dof$ reduced from 0.36 to 0.17, at the expense of errors in the lowest moments being increased by a factor between 2 or 3.
Central values and errors of $f_0(0)$, $|V_{cb}|\times 10^3$, ${\cal A}_0^0$ and ${\cal A}_0^+$ are now  $0.684 (18), 40.76 (70), 0.89 (35)$ and $1.93 (27)$, respectively. In addition, the higher moments are determined with large errors, ${\cal A}_1^0= 1.4 \pm 1.3 $ and ${\cal A}_1^+= 2.3 \pm 1.1$, and highly anticorrelated  ($r_{ij}\sim -0.91$) with their respective zero-order counterparts, as expected. Besides, the rest of the parameters are mostly statistically independent, except for a moderate $|V_{cb}|-f_0(0)$ correlation ($-0.63$) that still remains.  Thus, we conclude that the four-parameter fit of Table~\ref{tab:fit.combined} is quite robust in front of the inclusion of the next-to-leading terms in the expansion of Eq.~\eqref{eq.new}, especially for  $|V_{cb}|$ and $f_0(0)$. With regard to the phase moments the situation is not as satisfactory, since the covered $q^2$ range and the precision of the available inputs  are not enough to fully disentangle the first and second order phase moments. Nevertheless, the new ${\cal A}_0^0$ (${\cal A}_0^+$) is within 1 (2) $\sigma$ from  the value obtained in the leading order fit, which makes us confident to conclude that this phase moment is around  1 (2), if not larger.

Finally, we should mention that LQCD correlation matrices are not taken into account in the fit of Table~\ref{fig.fit}. Their inclusion produces variations in the central values of the fitted parameters, well taken into account by the errors collected in the table. Actually, these are negligible for $f_0(0)$, and only around 3/4, 5/6 and 3/5 of the corresponding sigmas for ${\cal A}_0^0$, ${\cal A}_0^+$ and  $|V_{cb}|$, respectively. In addition, errors on  $f_0(0)$, ${\cal A}_0^0$ and ${\cal A}_0^+$  become around a factor 1/2 smaller, driven by the FL-MILC input, which exhibits an extraordinary precision.  

}

%

\section{Comparison}

For decades, the CLN parametrization~\cite{Caprini:1997mu} has been widely used. In the work of Ref.~\cite{Caprini:1997mu}, the ratio
\bea
\frac{S_1(\omega)}{V_1(\omega)}&= & \frac{(1+r)^2}{2r(1+\omega)}\frac{f_0(\omega)}{f_+(\omega)} \nonumber  \\
&=&A\big[1+B(\omega-\omega_0)+C(\omega-\omega_0)^2+\cdots\big] 
\label{Eq_CLN_ratio}
\eea
is reported as a series of $\omega$ expanded around some $\omega_0$, with $r=m_D/m_B$. The coefficients $A,B$ and $C$ were determined from available LQCD results at that time, HQET and sum-rule calculations and unitary constrains, and included leading short-distance and $1/m_Q$ corrections~\cite{Neubert:1992tg,Neubert:1992qq} as well. Given the above relations and our new parametrization of $f_{+,0}(q^2)$ in Eq.~\eqref{eq.new}, we obtain the HQET prediction of the difference between $\A_0^+$ and $\A_0^0$ as
\bea
[\mathcal{A}_0^+-\mathcal{A}_0^0]_{\rm HQET}=\frac{(1+r)^2}{1-2\omega_0 r+r^2}\ln\frac{(1+r)^2}{2rA(1+\omega_0)}~~
\label{Eq_delta_A}
\eea
by matching at $\omega=\omega_0$. In Ref.~\cite{Caprini:1997mu}, $A$ was given by expanding  the results for the ratio of Eq.~\eqref{Eq_CLN_ratio} for two different choices of $\omega_0$ (see Tables~A.1 and A.2 of that reference). For $\omega_0=1$, $A=1.0036$, while $A=1.0018$  for $\omega_0\simeq1.267$. These spread of values for $A$ leads to
\bea
\label{eq.SMdiff}
[\mathcal{A}_0^+-\mathcal{A}_0^0]_{\rm HQET}\simeq 1.05\sim1.12
\, .
\eea
As mentioned,  Eq.~\eqref{Eq_delta_A} was obtained only from the constant term in Eq.~\eqref{Eq_CLN_ratio}. As a further check, we have also found the above difference of phase moments by matching the  $(\omega-\omega_0)$ term 
\bea
[\mathcal{A}_0^+-\mathcal{A}_0^0]_{\rm HQET}=\frac{(1+r)^2}{2r}\frac{B(1+\omega_0)+1}{(1+\omega_0)}
\eea
which gives values  in { the $1.13\sim 1.28$ range, showing some  inconsistency with those obtained from the first term of the expansion carried out in Ref.~\cite{Caprini:1997mu}. This gives strong indications that higher order HQET corrections, neglected in the CLN parametrization, are sizable, in agreement with the conclusion in Refs.~\cite{Bigi:2016mdz,Bigi:2017njr,Grinstein:2017nlq}.

The difference $[\mathcal{A}_0^+-\mathcal{A}_0^0]$ can be also obtained from our results in Table~\ref{tab:fit.combined},
\bea
[\A_0^+-\A_0^0]_{\rm this~work}=1.22\,(3)\, ,\label{eq:compHQET}
\eea
where we have taken into account the large statistical  correlation between $\mathcal{A}_0^+$ and $\mathcal{A}_0^0$ to obtain the error above.
Our result is larger than the HQET prediction in Eq.~\eqref{eq.SMdiff}, but it could be accommodated within the range deduced from the linear term of the CLN expansion. Furthermore, uncertainties in Eq.~\eqref{eq:compHQET} are certainly larger because of the systematic error produced by neglecting second-order moments, as discussed above. }

%

\section{Further considerations}

As we stressed above, one of the advantages of the  parametrization proposed in this work is that the fitted phase moments may be used to learn details on the $\bar B \bar D$ dynamics. Let us focus on $\A_0^0$, and let us note that if  $\alpha^0(s)$ is replaced by the constant  $\pi$ in Eq.~\eqref{eq.moments}, the zeroth order $S$-wave phase moment would be 1 (taking $s_0=0$). In the elastic region, $\sqrt{s}< (m_{B^*}+m_{D^*})$, the phase $\alpha^0$ coincides with the $S$-wave $\bar B\bar D$ phase shift. Let us suppose that the integration in  Eq.~\eqref{eq.moments} is being dominated by phase-space regions close to threshold; then according to Levinson's theorem, it would be justified to replace $\alpha^0(s)$ by $\pi$ if there exists one, but only one,  $\bar B \bar D$ bound state. This scenario easily explains
a value for $A_0^0$ of 1. Moreover, since the best fit value is $1.38 (12)$, we might conjecture either the existence of two bound states or of one bound and one virtual state\footnote{{We refer to a virtual state as a pole that is not located on the first Riemann sheet, but that nevertheless strongly influences the scattering in the physical region. A well known example can be found in the isovector $^1S_0$ [$^{2S+1} L_J$] nucleon-nucleon wave. }}. We recall here that for an energy-independent interaction, which seems  a reasonable approach to describe low-energy $S-$wave $\bar B\bar D$ scattering, Levison's theorem establishes that $\delta(s_{\rm th})=n_b \pi$, with $n_b$ being the number of bound states of the potential\footnote{An $S$-wave bound state of zero binding energy gives a contribution of $\pi/2$ instead of $\pi$.}, and $\delta(\infty)=0$ \cite{GalindoPascual}. In the case of  two $\bar B \bar D$ bound states, we envisage a situation where the phase shift takes the value of $2\pi$ at threshold and after decreases  with $\sqrt{s}$ (positive scattering length),  providing an integrated value larger than one for $\A_0^0$. In the second case, one bound and one virtual state, the phase shift begins taking the value of $\pi$ at threshold, but it would grow in the vicinity of $s=s_{\rm th}$ (negative scattering length) to make possible the phase moment to reach magnitudes of around 1.4. 
We notice, however, that the above discussion might be altered by inelastic-channel effects that will induce energy dependent interactions.


\section{Summary}

In this article, we have proposed a new model-independent parametrization for the form factors in the semileptonic $\bar{B}\to D\ell\nu$ decays. 
It provides an excellent simultaneous reproduction of experimental measurements of the differential decay rate and the LQCD results for $f_+$ and $f_0$, leading to a quite accurate determination of $|V_{cb}|$. We also confirm  that the previous tension between the exclusive and inclusive determinations was mostly due to the use of the CLN parametrization. Furthermore, the fitted phase moments $\A_0^{0,+}$ provide valuable information to constrain  the $S$- and $P$-wave $\bar B \bar D$ interactions. Any model for them should be consistent with the determination of these parameters extracted here from the $\bar B \to D$ semileptonic decays.
As an example, we have given strong hints about the existence of at least one bound and one virtual $\bar B \bar D$ $S$-wave $0^+$ states, subject to uncertainties produced by potentially sizable inelastic effects. 
The same parametrization can be also employed to other $b\to c$ semileptonic processes such as $\bar B\to D^* \ell \bar \nu_\ell$ and $\Lambda_b\to \Lambda_c\ell \bar \nu_\ell$.

\acknowledgments
{We warmly thank B. Grinstein for clarifying comments.} This research has been supported in part by the Spanish Ministerio de Econom\'ia y Competitividad (MINECO) and the European Regional Development Fund (ERDF) under Contracts No. FIS2017-84038-C2-1-P and No. SEV-2014-0398, by the EU STRONG-2020 project under the program H2020-INFRAIA-2018-1, Grant No. 824093 by the National Natural Science Foundation of China (NSFC) and the Deutsche Forschungsgemeinschaft (DFG) through funds provided to the Sino-German CRC 110 ``Symmetries and the Emergence of Structure in QCD" (NSFC Grant No. 11621131001), by NSFC under Grants No. 11835015, No. 11905258, No. 11947302 and 11961141012, by and the Fundamental Research Funds for the Central Universities under No. 531118010379, by the Chinese
Academy of Sciences (CAS) under Grants No. QYZDB-SSW-SYS013 and No. XDPB09, and by
the CAS Center for Excellence in Particle Physics (CCEPP).

\bibliography{B2Dbib}

\end{document}